%% file: main.tex
\pdfoutput=1
\documentclass[conference,a4paper]{IEEEtran}
\usepackage{cite}
\usepackage[hyphens]{url}
\usepackage[hidelinks]{hyperref}
\hypersetup{breaklinks=true}
\input{packages.tex}

\ifCLASSINFOpdf
\else
\fi

\begin{document}

\newpage

%
\title{Towards Hardware Support for FPGA Resource Elasticity}

\author{

Ahsan Javed Awan$^{1}$, Fidan Aliyeva$^{1,2}$\\

\vspace{-0.4cm} \normalsize $^1$Ericsson Research,
$^2$KTH Royal Institute of Technology\\\\

ahsan.javed.awan@ericsson.com, fidan.aliyeva040717@gmail.com 

}

\maketitle

\thispagestyle{plain}
\pagestyle{plain}

\definecolor{dgreen}{rgb}{0.00, 0.75, 0.00}
\definecolor{dred}{rgb}{0.75, 0.00, 0.00}
\definecolor{dblue}{rgb}{0.00, 0.00, 0.75}
\newcommand{\ahsan}[1]{[{\color{blue}Ahsan: #1}]}
\newcommand{\fidan}[1]{[{\color{red}Fidan: #1}]}

\input{sections/abstract.tex}

\IEEEpeerreviewmaketitle

\input{sections/introduction.tex}
\input{sections/background.tex}

\input{sections/relatedwork.tex}

\input{sections/methodology.tex}

\input{sections/results.tex}

\input{sections/conclusion.tex}


\bibliographystyle{IEEEtran}
\bibliography{IEEEabrv,ref}

\input{sections/APPENDIX.tex}

\end{document}

%% file: packages.tex
\usepackage{comment}
\usepackage{graphicx}
\usepackage{verbatim}

\usepackage{siunitx}

\usepackage{multirow}

\usepackage{tikz}
\usepackage{tikzscale}
\usepackage{pgfplots}
\usepackage{float}
\usepackage{todonotes}

\usepackage[labelfont=bf]{caption}
\usepackage{subcaption}

\usepackage{amssymb}
\usepackage{pifont}
\usetikzlibrary{patterns,arrows}
\usepackage{color}
\usepackage{xcolor}
\usepackage{soul}

\makeatletter
\IEEEtriggercmd{\reset@font\normalfont\fontsize{7.95pt}{8.0pt}\selectfont}
\makeatother
\IEEEtriggeratref{1}

%% file: sections/abstract.tex
\begin{abstract}
FPGAs are increasingly being deployed in the cloud to accelerate diverse applications. They are to be shared among multiple tenants to improve the total cost of ownership. Partial reconfiguration technology enables multitenancy on FPGA  by partitioning it into regions, each hosting a specific application's accelerator.  However, the region's size can not be changed once they are defined, resulting in the underutilization of FPGA resources. This paper argues to divide the acceleration requirements of an application into multiple small computation modules. The devised FPGA shell can reconfigure the available PR regions with those modules and enable them to communicate with each other over Crossbar interconnect with the Wishbone bus interface. The implemented crossbar's reconfiguration ability allows to modify the number of allocated modules and allocated bandwidth to those modules while preventing any invalid communication request. The envisioned resource manager can increase or decrease the number of PR regions allocated to an application based on its acceleration requirements and PR regions' availability.

\end{abstract}

%% file: sections/introduction.tex
\section{Introduction}
\label{sec:introduction}

With the advent of 3D ICs based on Stacked Silicon Interconnect (SSI) technology, the FPGA manufacturers are offering devices that are very rich in resources. Dedicating one large FPGA to a single user leads to poor utilization of the FPGA resources and a drastic increase in the infrastructure cost. To alleviate this, several proposals exist to support multi-tenancy on FPGAs, where FPGA’s internal resources are statically divided into multiple partitions~\cite{chen2014enabling,byma2014fpgas,fahmy2015virtualized,xia2016hypervisor,asiatici2017virtualized,zhu2018fpga,knodel2018fpgas,al2019cloud,zhang2017feniks,vaishnav2018resource}, each of which can be dynamically re-configured with the bitstreams using partial reconfiguration technology. The partial bitstreams of the requested accelerator are either made available by the framework in the database~\cite{chen2014enabling,fahmy2015virtualized,zhu2018fpga,knodel2018fpgas} or generated at runtime~\cite{byma2014fpgas,tarafdar2017designing,tarafdar2017heterogeneous,asiatici2017virtualized,al2019cloud,zhang2017feniks,vaishnav2018resource} by synthesizing the user-provided accelerator files described at RTL level, OpenCL or HLS level. The solutions target different CPU-FPGA architectures; i) PCIe attached FPGA~\cite{chen2014enabling,fahmy2015virtualized,tarafdar2017designing,tarafdar2017heterogeneous, asiatici2017virtualized,zhu2018fpga,knodel2018fpgas,zhang2017feniks}, ii) Network-attached FPGA~\cite{byma2014fpgas,weerasinghe2018standalone,al2019cloud,istvan2018providing} and iii) System on Chip comprising of ARM CPUs and programmable logic~\cite{vaishnav2018resource}. They provide isolation among multiple processes~\cite{fahmy2015virtualized,asiatici2017virtualized,zhang2017feniks}, OpenCL tasks~\cite{vaishnav2018resource} and virtual machines~\cite{chen2014enabling,byma2014fpgas,tarafdar2017designing,tarafdar2017heterogeneous,zhu2018fpga}, which are orchestrated via OpenStack.

Nevertheless, partially reconfigurable (PR) regions are fixed in size; it is not possible to dynamically increase or decrease the resources of PR once it is defined. This can lead to either underutilization of resources or not having PR regions large enough to host the application. In other words, the PR region might contain much more resources compared to the acceleration requirements of the application; thus, wasting some resources. On the other hand, it might also be the case that the application requires more resources than any PR region has; thus, not being able to host it. Therefore, the question is how to dynamically adjust PR regions accordingly to the application’s needs. While outlining our proposal, We make the following contributions;

\begin{itemize}

    \item We provide a system architecture that enables expanding and contracting the FPGA resources allocated to an application (also known as FPGA resource elasticity) by expressing application accelerators into multiple smaller computation modules. These modules are configured into multiple reconfigurable regions assigned to the application dynamically. Next, it provides a low-area-overhead, configurable, and isolated communication mechanism among those reconfigurable regions by adjusting the crossbar interconnect and WISHBONE (WB) interface~\cite{Peterson2001SpecificationFT}. 
    
    \item Compared to the prior-art~\cite{mbongue2020architecture}, our communication mechanism among PR regions is area efficient, i.e. consumes 61\% fewer lookup tables (LUTs) and 95\% fewer flip-flops (FFs). It also improves the latency to complete the request by 69\%.

\end{itemize}

%% file: sections/background.tex
\section{Background}
\label{sec:background}

\subsection{\textbf{Multiprocessor Interconnection Methods}}
Crossbar and Network-on-Chips (NoCs) are more effective ways of interconnection among multiple processors or chips compared to the shared bus. Since only one processor can access the bus at a time, a shared bus results in limited bandwidth and increased latency.

\begin{figure}
     \centering
     \begin{subfigure}[t]{0.49\columnwidth}
         \centering
         \includegraphics[width=\textwidth]{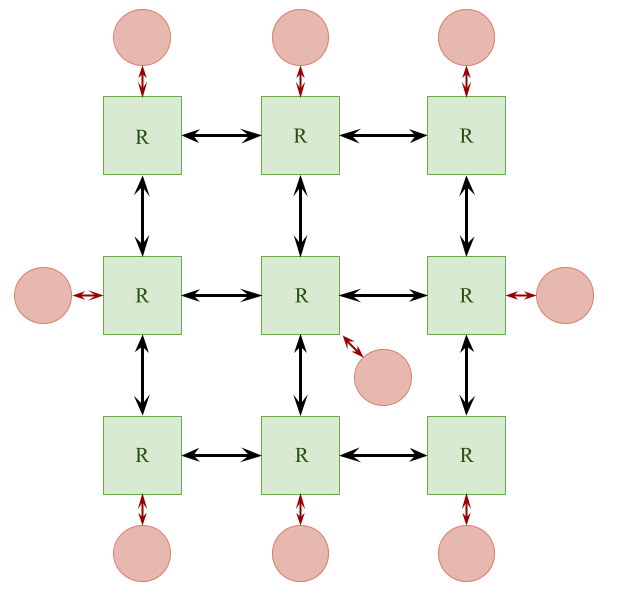}
         \caption{3x3 NoC with Mesh Topology}
         \label{fig:NoC}
     \end{subfigure}
     \hfill
     \begin{subfigure}[t]{0.49\columnwidth}
         \centering
         \includegraphics[width=\textwidth,height=4.25cm]{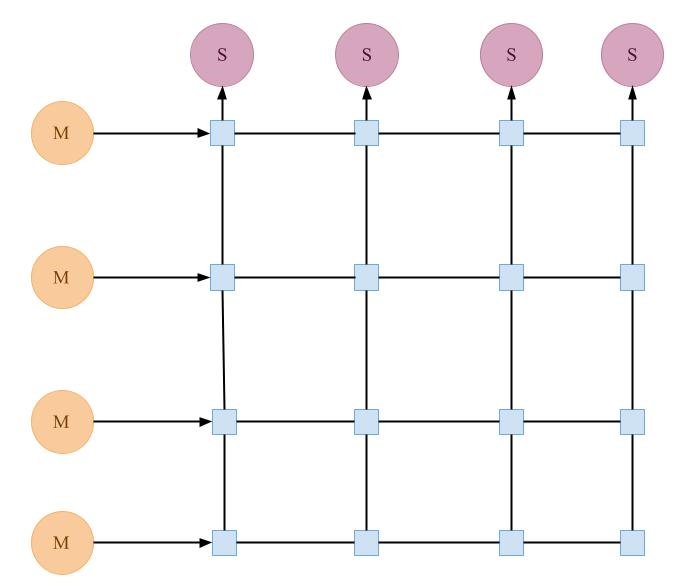}
         \caption{4x4 Crossbar Switch}
         \label{fig:Crossbar}
     \end{subfigure}
        \caption{Multiprocessor Interconnection on a Chip}
        \label{fig:noc_crossbar}
\end{figure}

\subsubsection{\textbf{NoCs}} 
Inspired by computer networks, in this scheme each module is considered as a node and has a router; then, routers are connected following one of the network topologies such as mesh (see Figure~\ref{fig:NoC}), torus, butterfly, and so on. To communicate, a node (represented in red) directs its data to its router (shown in green) and the router divides data into packets and forwards them into the next router. Thus, a packet travels through several routers until it reaches its destination. Routers can use either routing algorithms (random, weighted random, adaptive, etc.) or routing tables to decide which path to follow~\cite{dally2004principles}.
Since there are multiple paths and routers available in network topology, parallel transmissions can happen.

\subsubsection{\textbf{Crossbar Switch Interconnection}}
It consists of switches arranged in a matrix form and connects the source module to a destination. As displayed in Figure~\ref{fig:Crossbar}, all modules are connected to a common set of bus lines; however, their communications are prevented by switches illustrated as blue boxed. A master can initiate a request, upon which necessary switches are enabled to recover the physical link between a master and destination slave. Since there exist separate bus lines for each destination, in crossbar interconnection it is possible to do the parallel transmission. 

The crossbar switch is more flexible and scalable than the shared bus; however, less than NoCs. The flexibility and scalability result in increased resource usage for crossbar switch compared to shared bus~\cite{Peterson2001SpecificationFT,lee2004chip,lahtinen2003comparison} but it still consumes less area resource than NoCs. In other words, it stays in the middle of the trade-off between area and flexibility/scalability. Its area usage mainly comes from its arbitration logic~\cite{lahtinen2003comparison}. Crossbar can support parallel transmissions without incurring the protocol overhead as in NoCs. It can also demonstrate adequate latency and throughput in pipelined and global communication patterns~\cite{ryu2001comparison}.

\subsection{\textbf{WISHBONE Interconnection Architecture}}
It provides a standard data exchange protocol among IP cores in system-on-chip, connected in one of the supported communication interconnection modes (point-to-point, shared bus, data flow, and crossbar)~\cite{Peterson2001SpecificationFT}, thereby increasing the portability of IP cores. This protocol must be followed when designing WISHBONE bus interface. The interface operates with master-slave logic, where the master can initiate either read or write requests to a slave. For high throughput, a master sends all data if a slave has not asserted stall and then waits for acknowledgments from a slave.

WB interconnection has small design complexity and therefore, it can operate in high frequencies while consuming less area~\cite{Peterson2001SpecificationFT} and energy~\cite{lee2004chip}. Hardware modules can be adapted easily to interact with it, thereby increasing the reusability of the interconnection architecture. Moreover, it has a built-in handshaking protocol, eliminating the need for extra logic to ensure transmission safety.

%% file: sections/relatedwork.tex
\section{Related Work}
\label{sec:relatedwork}

Prior art on supporting resource elasticity inside the FPGAs~\cite{vaishnav2018resource} provides various implementations consuming different amounts of resources in the reconfigurable region for each application. Next, it selects kernel versions and the number of instances based on the fairness of resource allocation for each application and reconfigures the FPGA. It relies on bitstream manipulation to change the location of a module, increase/decrease allocated resources, etc. However, manipulating bitstreams is device-specific as different FPGA devices have different bitstream file formatting. 

Several communication schemes in the literature allow interaction among the partially reconfigurable regions. For example, a pipelined shared bus architecture with encapsulated WB (E-WB) interface for PR regions and 5-layer communication protocol~\cite{hagemeyer2007design}. Another technique~\cite{mbongue2020architecture} proposes NoCs among virtual regions. The network offers bufferless routers with no virtual channels. However, each virtual region contains an additional access monitor, wrapper, registers, and the communication interface with buffers beside the partially reconfigurable module. A different strategy explored in~\cite{ahmadinia2005practical} extends reconfigurable multiple buses on-chip for NoCs to overcome extra communication overhead. It provides a physical communication link between a source and destination of NoCs’ nodes. The link is divided into multiple bus segments, controlled by bus controllers deployed for each section. Likewise, physical channels are established over the crossbar switch before data transfers among the PR modules. Each connection point in Figure~\ref{fig:Crossbar} is implemented using one Configurable Logic Block (CLB)~\cite{fischer2010fpga}. These cross-points can be removed or recovered based on a user request. A control logic enables/disables the CLBs accordingly via a configuration signal. The setup channel provides a multiple 7-bit sections between the source and destination modules, limiting the number of source modules being able to communicate with a destination module.

Various studies focus on providing crossbar switch as a communication method for Multiprocessor System-on-Chip (MPSoC). For example, a configurable crossbar switch\cite{hong2006configurable} tends to reduce communication bottlenecks and improves the performance of the target system. Another reconfigurable crossbar switch~\cite{kim2005reconfigurable} has an additional communication line to allocate extra bandwidth for an IP.

The studies aforementioned, either based on NoCs~\cite{dally2004principles,bobda2005dynamic,pionteck2006applying} or shared bus ~\cite{hagemeyer2007design} can potentially expand or contract the FPGA resources allocated to application. Nevertheless, those architectures have several disadvantages. The shared bus method is neither flexible in supporting different communication patterns and parallel transmissions nor scalable to the higher number of modules due to its bandwidth limitations. On the contrary, NoC's large network protocol overhead, area usage, and power consumption are disadvantageous~\cite{lee2004chip,mak2006fpga}. The area-efficient crossbar switch solution~\cite{fischer2010fpga} is very low-level and less flexible. The physical channel can not be shared among multiple applications and limits the number of source modules for a destination by the number of available wire-bundles.

%% file: sections/methodology.tex
\section{Our Design}
\label{sec:methodology}
\begin{figure}[ht]
    \centering
    \includegraphics[width=0.8\columnwidth]{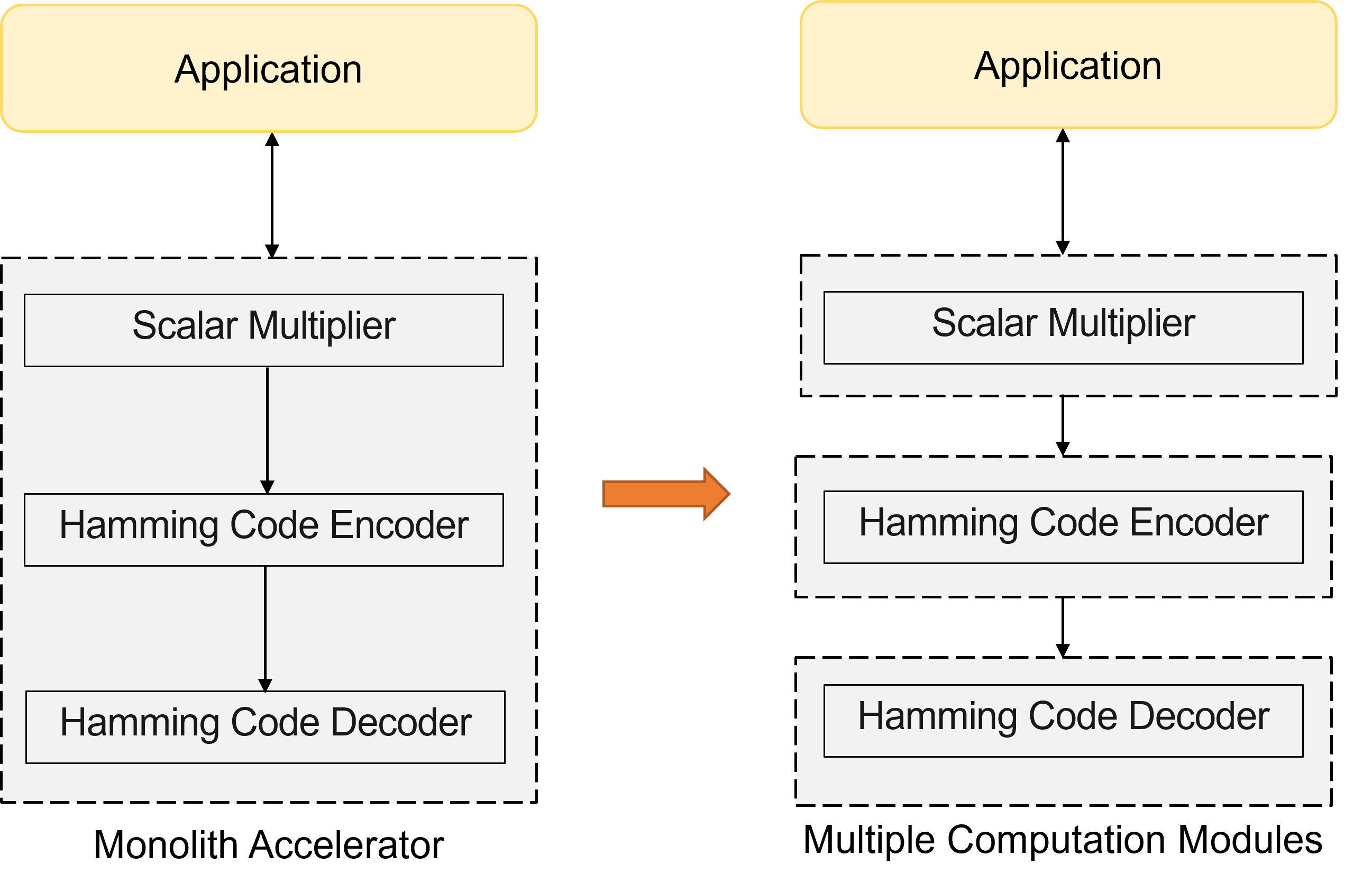}
   \caption{\emph{User's acceleration requirement expressed in small computation modules}
    \label{fig:accelerator-decompose}}
\end{figure}

\begin{figure}[ht]
    \centering
    \includegraphics[width=\columnwidth]{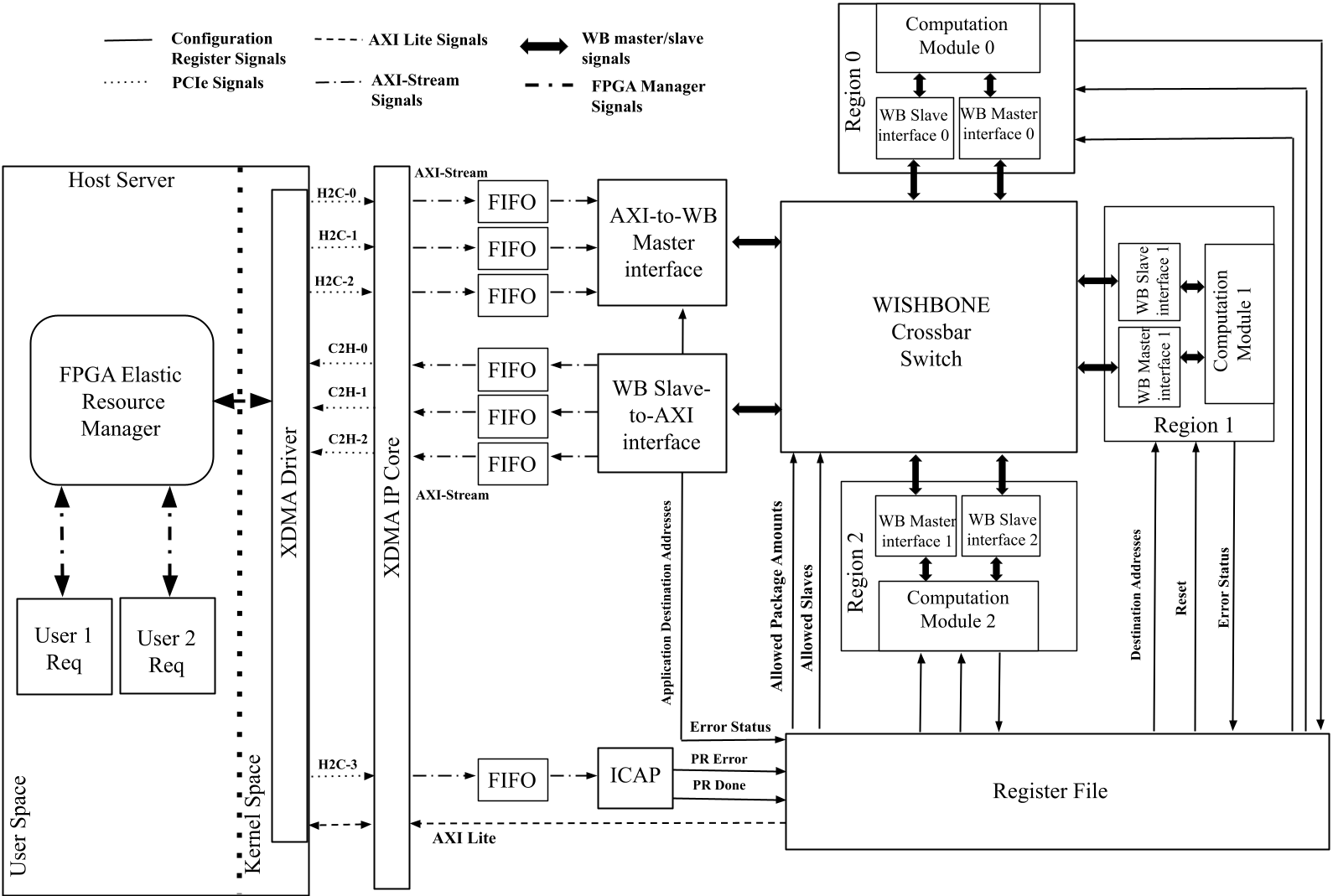}
   \caption{\emph{Overall System Architecture}
    \label{fig:bw_general_solution}}
\end{figure}

Our approach enables resource elasticity in the FPGA by providing effective communication method among PR regions. A user’s request for acceleration is expressed in the form of small computational modules (see Figure~\ref{fig:accelerator-decompose}), which are partially reconfigured to small-sized regions. This would improve resource under-utilization. By allocating extra PR regions if needed and enabling them to communicate using WB Crossbar Switch interconnection (see Figure~\ref{fig:bw_general_solution}), the number of allocated resources to the application can be also increased. Techniques to decompose the user's acceleration requirements into small computation modules are outside the scope of this paper.

Our solution enables dynamically changing FPGA PR region resources allocated to the application. It provides an easy way of managing communication isolation for different user requests and provides a simpler way of handling dynamic bandwidth allocation inside the FPGA device. It has low area usage due to optimized implementation. It incurs low communication or protocol overhead and is flexible to be used with different hardware modules due to the usage of the standard interface.

\subsection{\textbf{FPGA Elastic Resource Manager:}}
User requests are sent to the FPGA Elastic Resource Manager which keeps track of PR regions that are available and the regions are allocated to specific user's application. The manager analyzes the user request in terms of required PR regions to handle it and then, reconfigure the particular regions in the FPGA accordingly. Furthermore, it also sends user data to PR regions, provides configuration information to and recovers status information from the register file. 

Here is how reconfiguration and elasticity are achieved. The manager allocates the available amount of PR regions to the application’s computation modules through the internal configuration access port (ICAP). if there are not enough PR regions to host all modules, the remaining ones run on the server (referred to as on-server modules). Then, the manager provides configuration data to PR regions and the crossbar such as allowed modules, destination modules, and allowed number of packages. In this phase, the last module’s destination address is sent back to the server for the purpose of receiving its results and continuing the computation on the server. Afterward, it sends user data to start the computation process. When the on-server module finishes its computation, the FPGA manager checks again if there are any PR regions released so that it can run the on-server module on the FPGA, as well. If so, then it reprograms the available PR region with the on-server module and updates the other module’s destination addresses so that they communicate with the newly available module, as well. Thus, resource elasticity is achieved; the allocated resource for the user is increased by the means of communication interconnection among PR regions.

\subsection{\textbf{XDMA IP Core \& ICAP:}} Since the AXI-ST interface allows using each channel of XDMA (Xilinx Direct Memory Access) IP core~\cite{xilinxdma} separately, the design dedicates a separate channel to continuously stream partial bitstreams over the PCIe bus to saturate ICAP bandwidth~\cite{fastreconfig}. Moreover, FIFO is added before the ICAP to prevent data loss due to a mismatch in the clock frequency of ICAP (125 MHz) and of the rest of the system (250 MHz). Likewise, a separate AXI-Lite bypass link is enabled to access the register file to avoid interference between users' application data and configuration information.

\subsection{\textbf{Reset System}} 
Global reset is provided by buffering asynchronous reset signal of XDMA IP core. On the other hand, resets for computation modules and their associated crossbar ports are fed from the register file, thus during the partial reconfiguration process, the module can be isolated from the rest of the system and the crossbar port would be prevented from making any grant decisions.

\subsection{\textbf{Register File:}} 
The register file plays an important role in providing configuration data and storing necessary status information. Configuration data consists of the number of packages each module can send to each other and the destination address of each module. Firstly, when ICAP does reconfiguration of PR regions it stores the status data on the register file regarding whether the reconfiguration process was successful or failed.  Secondly, the WB Crossbar switch and PR regions are served by the register file, too. For the crossbar, the purpose of having the register file is to provide the allowed bandwidth to applications and to enable communication isolation. Moreover, error codes marking communication failure due to either wrong destination address or timeout due to unresponsive destination are also registered. Finally, reset signals for PR modules are provided through the register file too.

\subsection{\textbf{Crossbar Switch Architecture}} 
\begin{figure}
    \centering
    \includegraphics[scale=0.20]{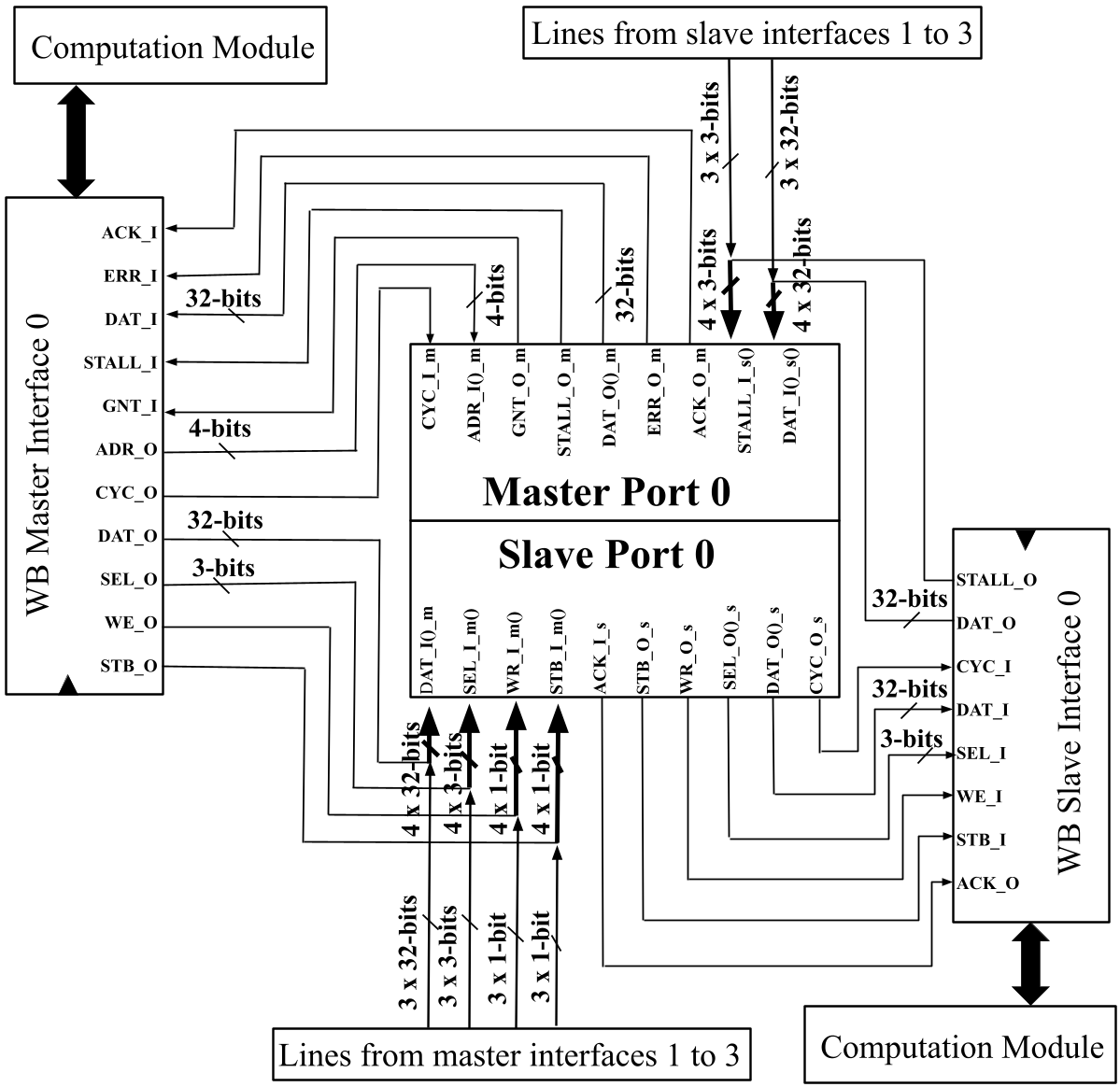}
   \caption{\emph{Block Diagram of the Proposed Crossbar Switch Interconnection }
    \label{fig:bw_port_0}}
\end{figure}

Figure~\ref{fig:bw_general_solution} shows each crossbar port consists of 2 different parts; these are called master and slave ports accordingly. The block diagrams of those ports are displayed in Figure~\ref{fig:bw_port_0} considering four-by-four crossbar interconnection.

\subsubsection{\textbf{Slave Port}} 
A slave port is responsible for giving grants based on requests coming from master ports. It also keeps the track of exchanged package numbers between a slave and a master. Additionally, it informs a master about the given grant and enables a slave for communication. This is done via an arbiter in each slave port serving masters, making the arbitration logic in this crossbar architecture decentralized. Finally, it connects granted master’s data signals to a slave interface through multiplexers. 

\textbf{Arbitration Logic – Weighted Round Robin:} To support bandwidth requirements of different accelerators, we propose Weighted Round Robin (WRR) arbiter based on leading zero counters (LZC)~\cite{oklobdzija1994algorithmic}, which operates at higher frequencies and has less area overhead~\cite{dimitrakopoulos2011scalable} compared to priority encoders based arbitration logic.

The arbiter ensures the customized bandwidth allocation. It tracks the number of packages rather than the time period via package counter, which looks up the registers holding the maximum number of packages each master is allowed to send. When the maximum number of packages is reached, it switches the grant to the next master . 
Having a decentralized arbitration scheme simplifies the arbiter logic and management of multicast data transmission.

\subsubsection{\textbf{Master Port}} 
It receives a communication request from a master interface together with the destination slave’s address. If a destination address is invalid it prevents the communication, returning an error signal. Otherwise, it directs a request to a slave port and waits for a grant. If a grant is given, it connects a target slave’s data lines to a master interface through multiplexers.

\textbf{Communication Isolation:} It is done with the help of configuration registers which provide a master port with allowed slaves. It has high bits for allowed slaves’ bit number while low bits for unallowed ones. Slave addresses are sent in one-hot encoding form by a master; for instance, to access slave 1, “0010” is sent. This eases the communication isolation as sent slave addresses and allowed addresses are compared with AND operator; if the result is 0, it means a master has sent an invalid slave address. In that case, the input port sends an error signal to a master and does not issue any request to a slave.

This method reduces the overhead of the configuration process because configuring a new module would require only updating the registers serving this module and the modules belonging to the same application. Moreover, validating the address on the slave side would incur extra clock cycles (ccs) from the arbiter logic on making a grant decision on an invalid master and then on generating an error signal.

\subsection{\textbf{WB Interfaces\iffalse and Computation Templates\fi}}
Next, we discuss the modified WB features desired to meet the needs of elastic resource management at reduced area overhead. 

\subsubsection{\textbf{ WB Master Interface}} 
Firstly, it initiates the request and provides the destination address to the crossbar upon receiving the request signal from a module, and then, it starts its watchdog timers. If it receives an error signal from the master port due to an invalid destination address or if the waiting time for a grant signal times out, it provides the error code back to a module. If a master is granted access to a slave, it issues data words together with their register addresses. Next in order, if the slave cannot serve the request currently; the master interface stops transmission and waits for the slave to become available. However, if the destination slave does not respond in a defined period, a timeout error happens. Otherwise, when all data is sent successfully it waits for acknowledgment signals for all transmitted data sets. 

\subsubsection{\textbf{ WB Slave Interface}} 
Upon receiving a valid request from a master, the slave interface enables its registers to store incoming data provided those registers currently do not contain any unread data, and sends an acknowledgment to a master. When registers become full, and a master still wants to send data slave interface stalls and de-asserts the acknowledgment informing a master that it needs to wait before sending new data and disables its registers. Meanwhile, it informs the computation module that its data buffer is full and waits for the module to read the data. The module triggers the slave interface once it has read the data, which causes the slave interface to reset its registers and start registering new data. Whenever the request is de-asserted, the slave interface goes into idle mode indicating either the master has no more data to send or it has sent the allowed number of packages by WRR or the master gives timeout error waiting long enough for the acknowledgment from a stalled slave.

\subsection{\textbf{AXI-to-WB and WB-to-AXI:}} Together with one of the crossbar's ports, these modules transfer data between computation modules and the user application. User data tagged with an application ID is received via any of the 3 host-to-card channels and stored in their FIFO buffers, each with an AXI interface. WB master interface in AXI-to-WB module serves each FIFO periodically through AXI interface. It looks up the ID in the register file, extracts destination modules, and delivers data to the destined PR region.  This prevents other applications to access unallocated PR regions even though the crossbar port has access to any PR region. Moreover, a master initiates a request as soon as the AXI side buffer becomes half full. It receives one 32-bit data word from FIFOs each cycle taking it 8 clock cycles to receive complete user data. Master also delivers 1 data word each clock cycle to the slave upon receiving the grant at best after 3 clock cycles. Therefore, overlapping 3 clock cycles of grant latency and 1 clock cycle of sending first data word with the second half of buffer receiving data from AXI end, the latency to deliver user data from FIFO to a computation module is reduced to 15 clock cycles compared to 19 clock cycles for the case where  AXI side buffer becomes full for a master to send request

Similarly, computation results can be read from any of the 3 card-to-host channels. WB slave interface in WB-to-AXI module sends these results to one of card-to-host channels through AXI-streaming interface. The selection of the interface is done based on the shift register which has 3 bits and only 1 bit enabled at a time. Consequently, each channel is targeted in a round-robin fashion.

\subsection{\textbf{Computation Module Template}} 
We provide a standard template for the computation modules to have the same interfaces. However, depending on application requirements or the nature of computation modules, the implementation can be different, which means interfaces need to be adapted to be operable with WB. Our standard template comprises input and output registers, error status register, computation units, and control logic. 

Upon receiving the buffer full signal from a slave interface, the control logic saves incoming data to input registers and signals the slave interface to register further incoming data. Since the first data word here indicates application ID, it is directly forwarded to the output register. Next, it enables the output registers to store the output of multiple computation units operating in parallel on the input data. Once the output is ready, it requests the master interface with output results and destination address. The status of the request is stored in the error register. If the request is successful, the output registers are reset. If a slave interface has new data, it registers new data; otherwise, it becomes idle. Furthermore, the error status is forwarded to the register file; hence, FPGA elastic resource manager can see if the status of the last request is successful or not.

%% file: sections/results.tex
\section{Experimental Results and Discussion}
\label{sec:results}

\subsection{\textbf{Tools}}
We use Xilinx KCU1500~\cite{kcu1500} acceleration development board to test our prototype. The board contains a Kintex Ultrascale XCKU115 FPGA device and is connected with a PCIe Gen3 to the server running Ubuntu 20.4 and XDMA drivers 2020.1.8~\cite{xdmadriver}.  We rely on Xilinx Vivado tool version 2018.3 to implement system architecture and describe different components in VHDL. 

\subsection{\textbf{System Implementation}}

The proposed system architecture contains an ICAP module to enable partial reconfiguration feature of the FPGA device; however, it has not been implemented in the current prototype as the overhead of partial reconfiguration on the same FPGA device has been covered in our earlier work~\cite{awanfpgakubernetes}.  Instead, the features of the proposed 32-bit WB Crossbar interconnect, are tested using statically allocated modules. The data width of WB Crossbar interface is chosen for fair comparison with related work e.g.~\cite{mbongue2020architecture}. 

The features of the implemented system architecture are summarized here; 1) Configurable 4 port WISHBONE Crossbar communication interconnection, which enables to increase or decrease the number of resources allocated to an application, allows dynamic bandwidth allocation for different applications, and provides communication isolation, 2) XDMA IP Core with 6 AXI-ST channels to exchange user data, 3) Three different statically implemented computation modules; the multiplier, the hamming encoder, and the hamming decoder together with WISHBONE master and slave interfaces, 4) AXI-WB and WB-AXI modules and 5) Register file to serve computation modules and the crossbar with configuration data and to save status data from computation and AXI-WB modules.

\subsection{\textbf{Resource Elasticity}}

\begin{figure}
    \centering
    \includegraphics[width=8cm]{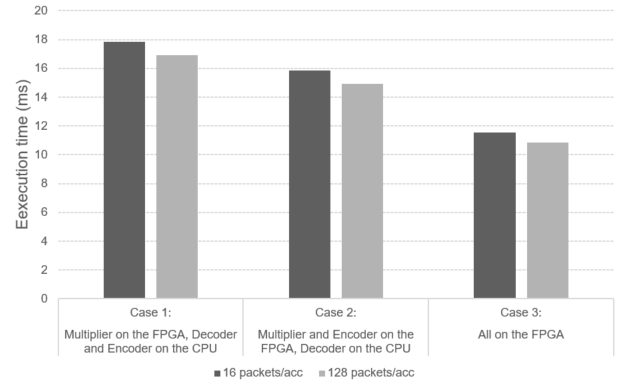}
   \caption{\emph{Comparison of Execution Time }
    \label{fig:elasticity_execution_time}}
\end{figure}

To show how elasticity improves the execution time of the application, we consider a use-case where 16 KB data is sent to be processed by the constant multiplier, the  Hamming code (31, 26) encoder, and decoder sequentially. We compare three cases; 
1) Multiplication is done on the FPGA while the CPU runs the rest.
2) The encoder becomes available; multiplier and encoder run on the FPGA while the decoder is on the CPU.
3) The decoder becomes available, as well, having all computations running on the FPGA.
For each case, the experiment is repeated 10 times and the average execution time is reported in Figure~\ref{fig:elasticity_execution_time}. While the average execution time is 16.9 ms for the 1$^{st}$ case, as the user gets more resources it improves to 10.87 ms due to resource elasticity.

\subsection{\textbf{Dynamic Bandwidth Allocation}}
To show the effect of dynamically configuring the WB Crossbar bandwidth allocated to computation modules, we consider the three cases mentioned above and repeat experiments at 16 and 128 packets, each packet of size 4 bytes, per accelerator by updating the register file and reporting the total execution time in the respective cases. One can see that by increasing the number of packets allocated to accelerators, the execution time improves from 5.24\% when one accelerator (multiply) is configured in the FPGA to 6\% when all three accelerators are configured on FPGA to communicate with each other over the WB Crossbar.

\subsection{\textbf{Communication Overhead}}
Time-to-grant is the number of clock cycles from the time when a computation module initiates a request to the time when a master interface starts to send the first data. It is 4 ccs in the best case, where the slave does not serve any request concurrently. It takes 2 ccs for the module’s request to reach the master interface and for it to initiate a request. Then, an arbiter spends 2 ccs to grant the request and enable the slave interface. If a computation module has 8 packages to deliver, the request completion latency is therefore 13 ccs. Here, the last clock cycle is used to register the error status of the transaction.
Along with this, the worst-case time-to-grant occurs when all 3 computation modules target the fourth one at the same time. The master being served the last would have to wait for the first 2 masters to be served. In this case, 13$^{th}$ cc for each previous master module can be ignored because a master interface releases the bus as soon as it completes sending its packages; thus, registering error code only on a master side. Consequently, the last computation module time-to-grant would be 28 ccs (12 ccs for each previous master and 4 ccs for time-to-grant) and request completion latency would be 37 ccs. 

\subsection{\textbf{Area \& Power Usage}}
As reported in Table~\ref{tab:area_all_comp}, the overall LUT and FF usage of the WB Crossbar together with computation modules and bus interfaces is 0.19\% and 0.1\% of KCU 1500’s LUT and FF resources. The WB crossbar interconnection itself uses 0.07\% LUTs and 0.004\% FFs. Depending upon the computation module, on average master and slave interfaces have 196 and 85 LUTs respectively, and correspondingly 117 and 628 FFs. . In addition, the LUT usage of the whole system architecture is 5.46\%  whose 5.04\% comes from XDMA IP Core. Correspondingly, the overall FF utilization is 2.75\% where XDMA IP contributes 2.32\%. WB Crossbar interconnection and a single master interface consume 1 mW power individually. However, a slave interface consumes less than 1mW. FPGA system consumes 5.03W whose 44\% comes from GTH transceivers.

The register file, as shown in Table 1, is implemented using LUTs and FFs. Our current implementation uses 20 registers combined in one register file. For more detailed description of the registers, please, refer to Table 3 in APPENDIX

\begin{table}[]
\renewcommand{\arraystretch}{1.3}
\caption{Area Usage of All Components}
\label{tab:area_all_comp}
\centering
\resizebox{\columnwidth}{!}{
\begin{tabular}{l|cccccc}
\hline
\multicolumn{1}{c|}{\textbf{\begin{tabular}[c]{@{}c@{}}Component\end{tabular}}} & \textbf{\begin{tabular}[c]{@{}c@{}} LUT\end{tabular}} & \textbf{\begin{tabular}[c]{@{}c@{}}\%\end{tabular}} & \textbf{\begin{tabular}[c]{@{}c@{}}FF\end{tabular}} & \textbf{\begin{tabular}[c]{@{}c@{}}\%\end{tabular}} & \textbf{BRAM} & \textbf{\begin{tabular}[c]{@{}c@{}}\%\end{tabular}} \\ \hline
XDMA IP Core & 33441 & 5.04 & 30843 & 2.32 & 62 & 2.87 \\ \hline
WB Crossbar & 475 & 0.07 & 60 & 0.004 & 0 & 0 \\ \hline
WB Hamming Decoder & 432 & 0.07 & 646 & 0.05 & 0 & 0 \\ \hline
WB Master Interface & 213 & 0.03 & 27 & \textless{}0.01 & 0 & 0 \\
WB Slave Interface & 115 & 0.02 & 220 & 0.02 & 0 & 0 \\
Hamming Decoder & 104 & 0.02 & 399 & 0.03 & 0 & 0 \\ \hline
WB Hamming Encoder & 233 & 0.04 & 99 & 0.01 & 0 & 0 \\ \hline
WB Multiplier & 138 & 0.06 & 624 & 0.05 & 0 & 0 \\ \hline
AXI-WB–FIFO System & 975 & 0.15 & 1842 & 0.14 & 13.5 & 0.62 \\ \hline
WB-AXI-FIFO System & 389 & 0.06 & 2274 & 0.17 & 13.5 & 0.62 \\ \hline
Register File & 265 & 0.04 & 560 & 0.04 & 0 & 0 \\ \hline
Total & 36348 & 5.47 & 36948 & 2.79 & 89 & 4.12 \\ \hline
\end{tabular}
}
\end{table}

\subsection{\textbf{Discussion}}
First and foremost, the resulting 32-bit width configurable crossbar communication interconnection enables dynamic resource allocation, dynamic bandwidth configuration, and communication isolation. Moreover, it takes a very small area (see Table \ref{tab:comp_exist_work}), which is 475 LUTs, 60 FFs, and no BRAMs while consuming 1 mW power to connect 4 modules. This number varies between 305 and 495 LUTs for a single 32-bit router in ~\cite{mbongue2020architecture}. Correspondingly 4 3-ports routers in a 2x2 NoC~\cite{mbongue2020architecture} would occupy 1220 LUTs and 1240 FFs to serve the same number of modules (see Table \ref{tab:comp_exist_work}). Consequently, our implementation takes 61\% less LUTs and 95\% fewer FFs than its equivalent NoC architecture. It also consumes 80x less power. 

Scaling the area usage of a single master-slave E-WB shared bus architecture~\cite{hagemeyer2007design} by a factor of 4 and comparing it to our 4x4 crossbar interconnection with WB master and slave interfaces, we find that our solution occupies 48.6\% more LUT resources and 46.4\% fewer FFs  (see Table \ref{tab:comp_exist_work}). A high LUT usage percentage is expected since, in general, crossbars occupy more area than the shared bus.

Next, time-to-grant in our design varies from 4 ccs constant in the best case to 28 ccs in the worst case, which might change depending on the bandwidth allocated to other modules. The current number assumes each module sends 8 packets. Compared to~\cite{ahmadinia2005practical}, a single command is processed in 8 ccs in the best case, 2 commands need to be exchanged between modules to start a communication, making it 16 ccs for time-to-grant. 

Finally, our solution takes 69\% less ccs than NoC based design~\cite{mbongue2020architecture} to complete a request. Because, a network package contains a head flit, tail flit, and body flits~\cite{dally2004principles}. Sending 8 sets of data, as in our case, would require sending 10 flits. The first flit takes 2 ccs to pass from one router. Due to pipelining, the remaining flits would take 1 cc each. Thus, traversing the flits only in source and destination routers would take 22 ccs as opposed to 13 ccs in our case.

\begin{figure}
    \centering
    \includegraphics[width=8cm]{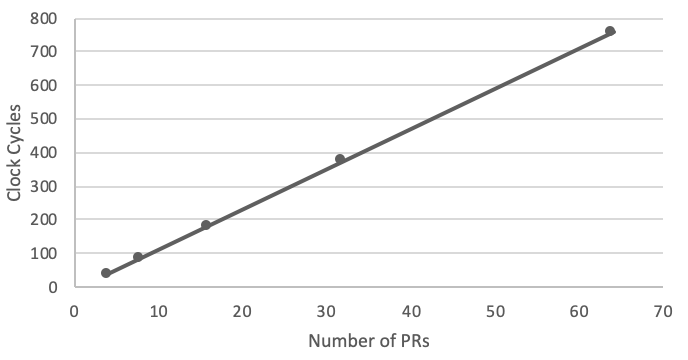}
   \caption{\emph{Number of PRs vs The Worst-Case Latency}
    \label{fig:PR_vs_ccs}}
\end{figure}

The area overhead of the LZC based arbiter increases quadratically with the number of ports. However, the increase rate is less than the other implementations~\cite{dimitrakopoulos2011scalable}. Increasing the partitions would require updating the register file with new registers, For each new coming PR region, three more registers has to be added; allowed addresses register, allowed package numbers register, and destination address register. Finally, the worst case latency increase would be linear as shown in Figure~\ref{fig:PR_vs_ccs}, where each master has eight data words to send.

\begin{table}[]
\renewcommand{\arraystretch}{1.3}
\caption{Comparison with Existing Work}
\label{tab:comp_exist_work}
\centering
\resizebox{\columnwidth}{!}{
\begin{tabular}{l|ccc}
\hline
\multicolumn{1}{c|}{\textbf{Design}} & \textbf{LUTs} & \multicolumn{1}{c}{\textbf{FFs}} & \textbf{Power (mW)} \\ \hline
4x4 WB Crossbar & 475 & 60 & 1 \\
2x2 NoC 3-port routers~\cite{mbongue2020architecture} & 1220 & 1240 & 80 \\
4x4 WB Crossbar Interconnection System & 1599 & 796 &  \\
4 Communication Infrastructures in~\cite{hagemeyer2007design} & 1076 & 1484 &  \\  \hline
\end{tabular}
}
\end{table}

%% file: sections/conclusion.tex
\section{Conclusion}
\label{sec:conclusion}
In this work, we have designed a low-area, low-communication overhead, configurable communication interconnection to enable FPGA resource elasticity. The resulting interconnection has intended features; supporting dynamically increasing or decreasing FPGA PR resources to an application while providing communication isolation. Moreover, dynamic bandwidth allocation to the application inside FPGA is one of the other important features of the resulting system. As expected, the resulting crossbar occupies much less area than NoCs and slightly more area than the shared bus. Future work includes assessing the overhead in detail when scaling our crossbar architecture and integrating the current implementation with PR technology and the Kubernetes engine to exploit the true potential of elasticity of FPGAs in the Cloud.

%% file: sections/APPENDIX.tex
\section{APPENDIX}
\label{sec:appendix}

\input{tables/register_file}

%% file: tables/register_file.tex
\begin{table}[ht]
\renewcommand{\arraystretch}{1.3}
\caption{Register File Description}
\label{tab:reg_file_desc}
\centering
\resizebox{\columnwidth}{!}{
\begin{tabular}{l|c|l}
\hline
{\textbf{N}} & {\textbf{Register Address}} & \textbf{Information Stored} \\ 
\hline
0  & 0x0              & FPGA device ID                                           \\
1  & 0x4              & PR region 1 destination address                        \\
2  & 0x8              & PR region 2 destination address                         \\
3  & 0xC              & PR region 3 destination address                         \\
4  & 0x10             & Reset PR regions and ports {[}3:0{]}                    \\
5  & 0x14             & Allowed Addresses of Port 0 Master                      \\
6  & 0x18             & Allowed Addresses of Port 1 Master                      \\
7  & 0x1C             & Allowed Addresses of Port 2 Master                      \\
8  & 0x20             & Allowed Addresses of Port 3 Master                      \\
9  & 0x24             & Package numbers allowed in port 0 for ports {[}3:0{]}   \\
10 & 0x28             & Package numbers allowed in port 1 for ports {[}3:0{]}   \\
11 & 0x2C             & Package numbers allowed in port 2 for ports {[}3:0{]}   \\
12 & 0x30             & Package numbers allowed in port 3 for ports {[}3:0{]}   \\
13 & 0x34             & Application ID 0 destination address                     \\
14 & 0x38             & Application ID 1 destination address                     \\
15 & 0x3C             & Application ID 2 destination address                     \\
16 & 0x40             & Application ID 3 destination address                     \\
17 & 0x44             & PR region {[}3:1{]} last transaction error status       \\
18 & 0x48             & App. ID {[}3:0{]} last transaction error status         \\
19 & 0x4C             & ICAP status                                            \\  \hline
\end{tabular}
}
\end{table}

%% file: main.bbl
\begin{thebibliography}{10}
\providecommand{\url}[1]{#1}
\csname url@samestyle\endcsname
\providecommand{\newblock}{\relax}
\providecommand{\bibinfo}[2]{#2}
\providecommand{\BIBentrySTDinterwordspacing}{\spaceskip=0pt\relax}
\providecommand{\BIBentryALTinterwordstretchfactor}{4}
\providecommand{\BIBentryALTinterwordspacing}{\spaceskip=\fontdimen2\font plus
\BIBentryALTinterwordstretchfactor\fontdimen3\font minus
  \fontdimen4\font\relax}
\providecommand{\BIBforeignlanguage}[2]{{%
\expandafter\ifx\csname l@#1\endcsname\relax
\typeout{** WARNING: IEEEtran.bst: No hyphenation pattern has been}%
\typeout{** loaded for the language `#1'. Using the pattern for}%
\typeout{** the default language instead.}%
\else
\language=\csname l@#1\endcsname
\fi
#2}}
\providecommand{\BIBdecl}{\relax}
\BIBdecl

\bibitem{chen2014enabling}
F.~Chen, Y.~Shan, Y.~Zhang, Y.~Wang, H.~Franke, X.~Chang, and K.~Wang,
  ``Enabling fpgas in the cloud,'' in \emph{Proceedings of the 11th ACM
  Conference on Computing Frontiers}, 2014, pp. 1--10.

\bibitem{byma2014fpgas}
S.~Byma, J.~G. Steffan, H.~Bannazadeh, A.~Leon-Garcia, and P.~Chow, ``Fpgas in
  the cloud: Booting virtualized hardware accelerators with openstack,'' in
  \emph{2014 IEEE 22nd Annual International Symposium on Field-Programmable
  Custom Computing Machines}.\hskip 1em plus 0.5em minus 0.4em\relax IEEE,
  2014, pp. 109--116.

\bibitem{fahmy2015virtualized}
S.~A. Fahmy, K.~Vipin, and S.~Shreejith, ``Virtualized fpga accelerators for
  efficient cloud computing,'' in \emph{2015 IEEE 7th International Conference
  on Cloud Computing Technology and Science (CloudCom)}.\hskip 1em plus 0.5em
  minus 0.4em\relax IEEE, 2015, pp. 430--435.

\bibitem{xia2016hypervisor}
T.~Xia, J.-C. Pr{\'e}votet, and F.~Nouvel, ``Hypervisor mechanisms to manage
  fpga reconfigurable accelerators,'' in \emph{2016 International Conference on
  Field-Programmable Technology (FPT)}.\hskip 1em plus 0.5em minus 0.4em\relax
  IEEE, 2016, pp. 44--52.

\bibitem{asiatici2017virtualized}
M.~Asiatici, N.~George, K.~Vipin, S.~A. Fahmy, and P.~Ienne, ``Virtualized
  execution runtime for fpga accelerators in the cloud,'' \emph{Ieee Access},
  vol.~5, pp. 1900--1910, 2017.

\bibitem{zhu2018fpga}
Z.~Zhu, A.~X. Liu, F.~Zhang, and F.~Chen, ``Fpga resource pooling in cloud
  computing,'' \emph{IEEE Transactions on Cloud Computing}, 2018.

\bibitem{knodel2018fpgas}
O.~Knodel, P.~R. Genssler, F.~Erxleben, and R.~G. Spallek, ``Fpgas and the
  cloud--an endless tale of virtualization, elasticity and efficiency,''
  \emph{International Journal on Advances in Systems and Measurements},
  vol.~11, no. 3-4, pp. 230--249, 2018.

\bibitem{al2019cloud}
A.~A. Al-Aghbari and M.~E. Elrabaa, ``Cloud-based fpga custom computing
  machines for streaming applications,'' \emph{Ieee Access}, vol.~7, pp.
  38\,009--38\,019, 2019.

\bibitem{zhang2017feniks}
J.~Zhang, Y.~Xiong, N.~Xu, R.~Shu, B.~Li, P.~Cheng, G.~Chen, and T.~Moscibroda,
  ``The feniks fpga operating system for cloud computing,'' in
  \emph{Proceedings of the 8th Asia-Pacific Workshop on Systems}, 2017, pp.
  1--7.

\bibitem{vaishnav2018resource}
A.~Vaishnav, K.~D. Pham, D.~Koch, and J.~Garside, ``Resource elastic
  virtualization for fpgas using opencl,'' in \emph{2018 28th International
  Conference on Field Programmable Logic and Applications (FPL)}.\hskip 1em
  plus 0.5em minus 0.4em\relax IEEE, 2018, pp. 111--1117.

\bibitem{tarafdar2017designing}
N.~Tarafdar, N.~Eskandari, T.~Lin, and P.~Chow, ``Designing for fpgas in the
  cloud,'' \emph{IEEE Design \& Test}, vol.~35, no.~1, pp. 23--29, 2017.

\bibitem{tarafdar2017heterogeneous}
N.~Tarafdar, T.~Lin, N.~Eskandari, D.~Lion, A.~Leon-Garcia, and P.~Chow,
  ``Heterogeneous virtualized network function framework for the data center,''
  in \emph{2017 27th International Conference on Field Programmable Logic and
  Applications (FPL)}.\hskip 1em plus 0.5em minus 0.4em\relax IEEE, 2017, pp.
  1--8.

\bibitem{weerasinghe2018standalone}
J.~Weerasinghe, ``Standalone disaggregated reconfigurable computing platforms
  in cloud data centers,'' Ph.D. dissertation, Technische Universit{\"a}t
  M{\"u}nchen, 2018.

\bibitem{istvan2018providing}
Z.~Istv{\'a}n, G.~Alonso, and A.~Singla, ``Providing multi-tenant services with
  fpgas: Case study on a key-value store,'' in \emph{2018 28th International
  Conference on Field Programmable Logic and Applications (FPL)}.\hskip 1em
  plus 0.5em minus 0.4em\relax IEEE, 2018, pp. 119--1195.

\bibitem{Peterson2001SpecificationFT}
W.~D. Peterson, ``{WISHBONE System-on-Chip (SoC) Interconnection Architecture
  for Portable IP Cores, B4 ed. OpenCores},'' 2010.

\bibitem{mbongue2020architecture}
J.~M. Mbongue, A.~Shuping, P.~Bhowmik, and C.~Bobda, ``Architecture support for
  fpga multi-tenancy in the cloud,'' in \emph{2020 IEEE 31st International
  Conference on Application-specific Systems, Architectures and Processors
  (ASAP)}.\hskip 1em plus 0.5em minus 0.4em\relax IEEE, 2020, pp. 125--132.

\bibitem{dally2004principles}
W.~J. Dally and B.~P. Towles, \emph{Routing Mechanics}.\hskip 1em plus 0.5em
  minus 0.4em\relax Elsevier, 2004, ch.~11, p. 203.

\bibitem{lee2004chip}
A.~S.-H. Lee and N.~W. Bergmann, ``On-chip interconnect schemes for
  reconfigurable system-on-chip,'' in \emph{Microelectronics: Design,
  Technology, and Packaging}, vol. 5274.\hskip 1em plus 0.5em minus 0.4em\relax
  International Society for Optics and Photonics, 2004, pp. 442--453.

\bibitem{lahtinen2003comparison}
V.~Lahtinen, E.~Salminen, K.~Kuusilinna, and T.~Hamalainen, ``Comparison of
  synthesized bus and crossbar interconnection architectures,'' in
  \emph{Proceedings of the 2003 International Symposium on Circuits and
  Systems, 2003. ISCAS'03.}, vol.~5.\hskip 1em plus 0.5em minus 0.4em\relax
  IEEE, 2003, pp. V--V.

\bibitem{ryu2001comparison}
K.~K. Ryu, E.~Shin, and V.~J. Mooney, ``A comparison of five different
  multiprocessor soc bus architectures,'' in \emph{Proceedings Euromicro
  Symposium on Digital Systems Design}.\hskip 1em plus 0.5em minus 0.4em\relax
  IEEE, 2001, pp. 202--209.

\bibitem{hagemeyer2007design}
J.~Hagemeyer, B.~Kettelhoit, M.~Koester, and M.~Porrmann, ``A design
  methodology for communication infrastructures on partially reconfigurable
  fpgas,'' in \emph{2007 International Conference on Field Programmable Logic
  and Applications}.\hskip 1em plus 0.5em minus 0.4em\relax IEEE, 2007, pp.
  331--338.

\bibitem{ahmadinia2005practical}
A.~Ahmadinia, C.~Bobda, J.~Ding, M.~Majer, J.~Teich, S.~P. Fekete, and J.~C.
  van~der Veen, ``A practical approach for circuit routing on dynamic
  reconfigurable devices,'' in \emph{16th IEEE International Workshop on Rapid
  System Prototyping (RSP'05)}.\hskip 1em plus 0.5em minus 0.4em\relax IEEE,
  2005, pp. 84--90.

\bibitem{fischer2010fpga}
T.~Fischer, ``Fpga crossbar switch architecture for partially reconfigurable
  systems,'' 2010.

\bibitem{hong2006configurable}
B.-J. Hong, K.-S. Cho, S.-H. Kang, S.-Y. Lee, and J.-D. Cho, ``On the
  configurable multiprocessor soc platform with crossbar switch,'' in
  \emph{APCCAS 2006 - 2006 IEEE Asia Pacific Conference on Circuits and
  Systems}.\hskip 1em plus 0.5em minus 0.4em\relax IEEE, 2006, pp. 1087--1090.

\bibitem{kim2005reconfigurable}
D.~Kim, K.~Lee, S.-j. Lee, and H.-J. Yoo, ``A reconfigurable crossbar switch
  with adaptive bandwidth control for networks-on-chip,'' in \emph{2005 IEEE
  International Symposium on Circuits and Systems (ISCAS)}.\hskip 1em plus
  0.5em minus 0.4em\relax IEEE, 2005, pp. 2369--2372.

\bibitem{bobda2005dynamic}
C.~Bobda and A.~Ahmadinia, ``Dynamic interconnection of reconfigurable modules
  on reconfigurable devices,'' \emph{IEEE Design \& Test of Computers},
  vol.~22, no.~5, pp. 443--451, 2005.

\bibitem{pionteck2006applying}
T.~Pionteck, R.~Koch, and C.~Albrecht, ``Applying partial reconfiguration to
  networks-on-chips,'' in \emph{2006 International Conference on Field
  Programmable Logic and Applications}.\hskip 1em plus 0.5em minus 0.4em\relax
  IEEE, 2006, pp. 1--6.

\bibitem{mak2006fpga}
T.~S. Mak, P.~Sedcole, P.~Y. Cheung, and W.~Luk, ``On-fpga communication
  architectures and design factors,'' in \emph{2006 International Conference on
  Field Programmable Logic and Applications}.\hskip 1em plus 0.5em minus
  0.4em\relax IEEE, 2006, pp. 1--8.

\bibitem{xilinxdma}
Xilinx, ``Dma/bridge subsystem for pci express v4.1,''
  \url{https://www.xilinx.com/support/documentation/ip_documentation/xdma/v4_1/pg195-pcie-dma.pdf},
  2021.

\bibitem{fastreconfig}
{Xilinx}, ``{Fast Partial Reconfiguration Over PCI Express, XAPP1338 (v1.0)},''
  \url{https://www.xilinx.com/support/documentation/application_notes/xapp1338-fast-partial-reconfiguration-pci-express.pdf},
  2019.

\bibitem{oklobdzija1994algorithmic}
V.~G. Oklobdzija, ``An algorithmic and novel design of a leading zero detector
  circuit: Comparison with logic synthesis,'' \emph{IEEE Transactions on Very
  Large Scale Integration (VLSI) Systems}, vol.~2, no.~1, pp. 124--128, 1994.

\bibitem{dimitrakopoulos2011scalable}
G.~Dimitrakopoulos, C.~Kachris, and E.~Kalligeros, ``Scalable arbiters and
  multiplexers for on-fgpa interconnection networks,'' in \emph{2011 21st
  International Conference on Field Programmable Logic and Applications}.\hskip
  1em plus 0.5em minus 0.4em\relax IEEE, 2011, pp. 90--96.

\bibitem{kcu1500}
Xilinx, ``{KCU1500 Board User Guide (UG1260)},''
  \url{https://www.xilinx.com/support/documentation/boards_and_kits/kcu1500/ug1260-kcu1500-data-center.pdf}.

\bibitem{xdmadriver}
------, ``{Xilinx DMA IP Reference drivers},''
  \url{https://github.com/Xilinx/dma_ip_drivers}.

\bibitem{awanfpgakubernetes}
A.~J. Awan, S.~Baig, and K.~Fertakis, ``{Data Communication Between a Host
  Computer and an FPGA},''
  \url{https://worldwide.espacenet.com/patent/search/family/069650683/publication/WO2021162591A1?q=pn\%3DWO2021162591A1},
  Aug.~19 2021, {WO2021162591A1}.

\end{thebibliography}
